\documentclass[english,a4paper,11pt]{article}

\usepackage[T1]{fontenc}
\usepackage{amssymb}
\usepackage[dvips]{graphicx}

\usepackage{hyperref}

\newtheorem{tm}{Theorem}

\newtheorem{cor}[tm]{Corollary}

\usepackage{soul}
\usepackage{xcolor}
\definecolor{cobalt}{rgb}{0.0, 0.28, 0.67}

\title{Some notes on the superintuitionistic logic of chequered subsets
  of $\mathbb{R}^\infty$}
\author{Tadeusz Litak}
\date{}

\begin{document}

\maketitle

{ \scriptsize\noindent
\textbf{Note 2018:} \textcolor{cobalt}{This paper was originally published in the Bulletin of the Section of Logic, vol. 33(2), pp. 81--86, 2004. This version (prepared soon after the paper was published and previously only available on my webpage) has been slightly extended; in particular, the proof of main theorem is hopefully more readable than in the Bulletin version. Very interesting follow-up results, settling in the negative questions posed in the final paragraph of this paper, have been announced by Ga\"{e}lle Fontaine and Timofei Shatrov. The former author shows in her AiML 2006 paper that {\sf ML} is not finitely axiomatizable over {\sf Cheq}.  The latter author claims to have settled negatively the issue of finite axiomatizably of {\sf Cheq} itself; however, to the best of my knowledge, this has never been published. Speaking of unpublished work, I also cannot find any final journal version of Grigolia's results \cite{grig} referred to in this paper. If his claims have never been verified, this also leaves open the status of the last sentence in the statement of Corollary \ref{cor:proml}.} 
}

\begin{abstract}\noindent
  We are going to investigate the superintuitionistic analogue of the
  modal logic of \textit{chequered} subsets of $\mathbb{R}^\infty$
  introduced by van Benthem et al.~\cite{vb} It will be observed
  that this logic possesses
  the disjunction property, contains the Scott axiom, fails to
  contain the
  Kreisel-Putnam axiom and is not structurally complete. We will
  prove that it is a sublogic of the Medvedev logic {\sf
  ML}. 
\end{abstract}

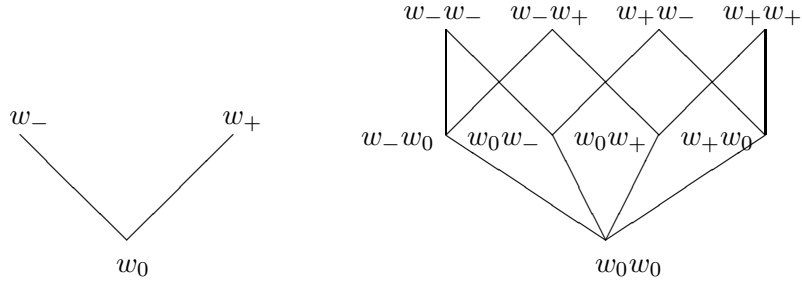
\begin{figure}[hbtp]
  \begin{center}
    \begin{picture}(90,40)
      
      \put(10,20){\line(1,-1){10}}
      \put(20,10){\line(1,1){10}}

      \put(9,21){$w_-$}
      \put(29,21){$w_+$}

      \put(19,7){$w_0$}

      \put(64,7){$w_0w_0$}

      \put(42,19){$w_-w_0$}
      \put(52,19){$w_0w_-$}
      \put(62,19){$w_0w_+$}
      \put(72,19){$w_+w_0$}

      \put(46,31){$w_-w_-$}
      \put(56,31){$w_-w_+$}
      \put(66,31){$w_+w_-$}
      \put(76,31){$w_+w_+$}

      \put(65,10){\line(-3,2){15}}
      \put(65,10){\line(-1,2){5}}
      \put(65,10){\line(1,2){5}}
      \put(65,10){\line(3,2){15}}

      \multiput(50,20)(30,0){2}{\line(0,1){10}}
      \multiput(50,20)(10,0){3}{\line(1,1){10}}
      \multiput(50,30)(10,0){3}{\line(1,-1){10}}

    \end{picture}
    \caption{Frames $\mathfrak{F}_1$ and $\mathfrak{F}_2$}
    \label{fig:f}
  \end{center}
\end{figure}

\noindent
In recent years, there seems to be growing interest in modal logics
determined by various topological spaces and particular families of
their subsets. Bezhanishvili et al.~\cite{bg} improved on a classical result by McKinsey and Tarski that
{\sf S4} is complete with respect to the real line by showing that
it is actually enough to consider only countable unions of convex
subsets. On the other hand, Aiello et al.~\cite{abb} proved that
the modal logic determined by finite unions of convex subsets of
$\mathbb{R}$ is a very strong tabular extension of {\sf Grz}
complete with respect to 2-fork Kripke frame $\mathfrak{F}_1$. $\mathfrak{F}_1 = \langle W_1,
R_1 \rangle$ is the first frame in Figure \ref{fig:f};  $W_1 = \{ w_0, w_-, w_+ \}$, all points are $R_1$-reflexive,
$w_0$ $R_1$-sees all the other points. Van Benthem et
al. \cite{vb} investigated logics determined by finite unions of
products of convex subsets of $\mathbb{R}$ in $\mathbb{R}^\alpha$ (where $\alpha \in
\mathbb{N} \cup \{ \infty \}$); such subsets were called
\textit{chequered}. It was established that for $\alpha = n$,  the modal logic in question corresponds to the logic
determined by $\mathfrak{F}_n  =
\mathfrak{F}^{n-1} \times \mathfrak{F}$, the order being the standard
product order. In case of $\alpha = \infty$, the respective modal
logic is determined by infinite sequence of frames $\{
\mathfrak{F}_n \}_{n \geq 1}$.

\noindent
It is, however, worth recalling that there exists another, simpler
tool well-tailored for describing topological spaces: it is the
language of
intuitionistic propositional logic and its extensions. In
particular, there is a strict correspondence between normal extensions of
{\sf Grz} and intermediate logics, as follows e.g., from Blok-Esakia
theorem and transfer results of Chagrov and Zakharyaschev (cf.,
e.g., Chagrov et al. \cite{chag}). Thus, the results of van Benthem et
al. \cite{vb} describe semantically superintuitionistic logics
determined by finite unions of products of open intervals. We will follow this line of
investigation, denoting the logic determined by $\{
\mathfrak{F}_n \}_{n \geq 1}$ as {\sf Cheq}. It will be proven
that this logic possesses the disjunction property, contains Scott axiom
$\mathsf{sa}$ and fails to contain the Kreisel-Putnam axiom
$\mathsf{kp}$.

\begin{figure}[hbtp]
  \begin{center}
    \begin{picture}(90,40)
      
      \put(10,20){\line(1,-1){10}}
      \put(20,10){\line(1,1){10}}

      \put(9,21){$\{1\}$}
      \put(29,21){$\{2\}$}

      \put(19,7){$\emptyset$}

      \put(64,7){$\emptyset$}

      \put(43,19){$\{1\}$}
      \put(58,19){$\{2\}$}
      \put(73,19){$\{3\}$}

      \put(46,31){$\{1,2\}$}
      \put(61,31){$\{1,3\}$}
      \put(76,31){$\{2,3\}$}

      \put(65,10){\line(-3,2){15}}
      \put(65,10){\line(0,1){10}}
      \put(65,10){\line(3,2){15}}

      \multiput(50,20)(30,0){2}{\line(0,1){10}}
      \multiput(50,20)(15,0){2}{\line(3,2){15}}
      \multiput(50,30)(15,0){2}{\line(3,-2){15}}

    \end{picture}
    \caption{Frames $\mathfrak{M}_1$ and $\mathfrak{M}_2$}
    \label{fig:m}
  \end{center}
\end{figure}
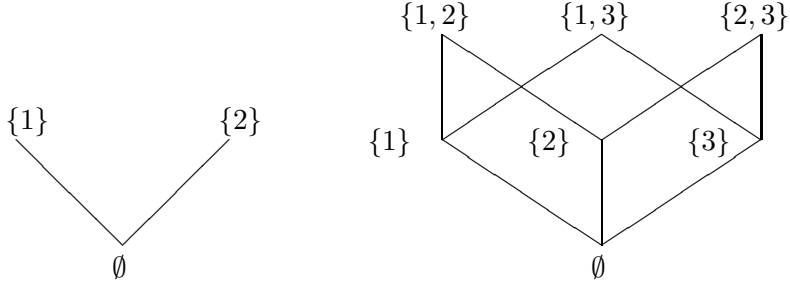

This perspective allows us to compare {\sf Cheq} with other
superintuitionistic systems. Perhaps the most famous semantically defined
one is Medvedev logic {\sf ML}. It is determined by the
class of all Boolean cubes with top element deleted --- i.e., by the
sequence of frames $\{\mathfrak{M}_n\}_{n \geq 1}$ where each
$\mathfrak{M}_n$ is the set of all proper subsets of $\{1, \ldots,
n+1\}$ ordered by the standard inclusion relation (Figure \ref{fig:m}). It is known from
Maksimova et al. \cite{maks} that
this logic is not finitely axiomatizable, its decidability being
still one of the most famous open problems in the field. As perhaps
the most interesting part of this short note, we are going to prove
that {\sf Cheq} is a proper sublogic of {\sf ML}.

\noindent
By $At(n)$ we shall denote the set of immediate successors of the
root of $\mathfrak{F}_n$.  For example, $At(1) = \{ w_-, w_+\}$,
$At(2) = \{ w_-w_0, w_+w_0, w_0w_-, w_0w_+\}$ and so on. The only coordinate at which $\bar{x}$
is distinct from $w_0$ will be denoted as $\uparrow\!\bar{x}$ i.e., for
any $\bar{x} \in At(n)$, $\uparrow\!\bar{x} = i$ iff $x_i \neq
w_0$. For example, $\uparrow\! w_0w_0w_- = 3$. By $_n|\bar{x}$ we will
denote $\bar{x}$ with $n$ leftmost coordinates deleted; analogously,
$\bar{x}|_n$ will denote $\bar{x}$ with $n$ rightmost coordinates
deleted. If $\bar{x}$ belongs to $\mathfrak{F}_m$, then both $_n|\bar{x}$
and $\bar{x}|_n$ belong to $\mathfrak{F}_{m-n}$.

\begin{tm} \label{dis}
{\sf Cheq} has the disjunction property.
\end{tm}

\noindent{\bf Proof}. Exactly as in the case of {\sf ML}; it is known that a
logic $L$ has the disjunction property iff it is characterized by a
class $\mathfrak{C}$ of descriptive rooted frames s.t. the disjoint
union of any two rooted frames from $\mathfrak{C}$ is a generated subframe
of a frame for $L$ (cf. Chagrov et al. \cite[Theorem 15.5]{chag}). It may be easily seen that
$\mathfrak{F}_n + \mathfrak{F}_m$ is a generated subframe of $\mathfrak{F}_{n
  + m}$. Indeed, one of the facts that make comparison of {\sf
  Cheq} and {\sf ML} interesting is that frames for both systems have the
property of {\em self-resemblance}; any rooted generated subframe of a
frame in the sequence is
isomorphic to some frame earlier in the list. This fact will be
used continually from now on. \hfill $\dashv$

\begin{tm}
{\sf Cheq} contains the Scott axiom 
\[
\mathsf{sa} = ((\neg\neg p
\rightarrow p) \rightarrow (p \vee \neg p)) \rightarrow (\neg p \vee
\neg\neg p)
\]
\end{tm}

\noindent{\bf Proof}. By induction; we are going to use the above-mentioned
self-resemblance property.
\begin{enumerate}
\item The Scott axiom holds in $\mathfrak{F}_1$.
\item Assume that the Scott axiom holds in $\mathfrak{F}_1, \ldots,
  \mathfrak{F}_n$. Hence, (by the self-resemblance property) the
  only point of $\mathfrak{F}_{n+1}$ where it can fail under some
  valuation $\mathfrak{V}$ is the root 
  $\bar{r} = w_0\ldots w_0$. We get that
\begin{eqnarray}
r & \vDash_\mathfrak{V} & (\neg\neg p \rightarrow p) \rightarrow (p
\vee \neg p), \label{ant} \\
r & \nvDash_\mathfrak{V} & \neg p \vee \neg\neg p \quad \mbox{but}
\label{cons_1} \\
x & \vDash_\mathfrak{V} &  \neg p \vee \neg\neg p \quad \mbox{for
  every other~} x. \label{cons_2}
\end{eqnarray}
Statement \ref{cons_2} is obtained from \ref{ant} by the
above-mentioned fact that no proper successor of the root can fail
the Scott axiom. This statement implies that any immediate successor of $r$
(i.e. each element of $At(n+1)$)
satisfies either $\neg p$ or $\neg\neg p$ under $\mathfrak{V}$. If all of them satisfy
$\neg p$, then it would contradict \ref{cons_1}; for the same
reason, at least one point in $At(n+1)$ must refute $\neg\neg
p$ (and hence verify $\neg p$). Hence, we have that there are $s, s' \in At(n+1)$ s.t. $s
\vDash_\mathfrak{V} \neg\neg p$ and $s' \vDash_\mathfrak{V} \neg
p$. Every maximal successor of $s$ must then verify $\neg p$ and every
maximal successor of $s'$ must verify $p$. But now the
following fact holds:
\begin{center}
For any $u, u' \in At(n+1)$, there exist $v(u, u') \in At(n+1)$
s.t. $v(u, u')$
has a common (maximal) successor both with $u$ and $u'$.
\end{center}
Hence, $v(s, s')$ fails $\neg p \vee \neg\neg p$, a contradiction
with \ref{cons_2}. \hfill $\dashv$

\end{enumerate}

\begin{tm} \label{kp}
{\sf Cheq} does not contain the Kreisel-Putnam axiom 
\[
\mathsf{kp} = (\neg p \rightarrow q \vee r) \rightarrow (\neg p
\rightarrow q) \vee (\neg p \rightarrow r)
\]
\end{tm}

\noindent{\bf Proof}. It fails in frame $\mathfrak{F}_2$ under valuation
$\mathfrak{V}$ defined as follows:
\[
\mathfrak{V}(p) := \{w_-w_+, w_+w_-\}, \quad \mathfrak{V}(q) :=
\{w_-w_-\}, \quad \mathfrak{V}(r) := \{ w_+w_+ \}. 
\]
\ \hfill $\dashv$

\begin{tm} \label{pmorph}
For every $n \geq 0$, $\mathfrak{M}_n$ is a (generated subframe of
a) p-morphic image of some $\mathfrak{F}_k$.
\end{tm}

\noindent{\bf Proof}. We are going to prove it by showing that for every \\ $n =
2^m-1~(m \geq 1)$ there exists a
p-morphism $f_n$ from $\mathfrak{F}_n$ onto $\mathfrak{M}_n$. The
restriction on $n$ is motivated only by
reasons of compactness and convenience; it is possible to prove
Theorem \ref{pmorph}
for any $n$.

\begin{enumerate}
\item For $m=1$, $f_1$ is just an isomorphism.
\item Assume that $f_{2^m - 1}$ is defined, $p = m+1$, $n = 2^p-1$.
\[
f_n(\bar{x}) = \left\{ \begin{array}{r@{\quad:\quad}l}
f_{2^m - 1}(\bar{x}|_{2^m}) & \bar{x} \in At(n), \uparrow\!\bar{x} <
2^m \\
\{ k + 2^m~|~k \in f_{2^m - 1}(_{2^m-1}|\bar{x}|_1)\} & \bar{x} \in At(n),
2^m \leq \uparrow\!{\bar{x}} < n \\
\{1, \ldots, 2^m \} & \bar{x} \in At(n), x_n = w_-
\\
\{2^m+1, \ldots, 2^p\} & \bar{x} \in At(n), x_n =
w_+ \\
\bigcup\{f_n(\bar{y})~|~\bar{y} \in At(n), \bar{y} R_n
\bar{x}\} & \bar{x} \notin At(n)
\end{array}
\right.
\]
\end{enumerate}

\noindent
We will sketch why this is a p-morphism onto
$\mathfrak{M}_n$. The proof is by induction; assume that $m>1$ and $f_{2^m-1}$
is a p-morphism.
\begin{itemize}
\item The image of $\mathfrak{F}_n$ via $f_n$, $f_n[\mathfrak{F}_n]$,
  is contained in $\mathfrak{M}_n$. As by definition $f_n(\bar{y})$ is a
  subset of $\{1, \ldots, 2^p\}$ for any $\bar{y}$, we only have to show
  that for no $\bar{y}$, $f_n(\bar{y}) = \{1, \ldots, 2^p\}$.  Assume
  that $f_n(\bar{y}) \supseteq \{1, \ldots, 2^m\}$. 
  $f_n(\bar{y}) = \bigcup\{f_n(\bar{x})~|~\bar{x} \in At(n), \bar{x} R_n
\bar{y}\}$. For  $\bar{x} \in At(n)$, $f_n(\bar{x}) \cap \{1, \ldots,
2^m\} \neq \emptyset$ only if $\uparrow\bar{x}<2^m$ or
$\uparrow\bar{x}=n$. But by the induction assumption $f_{2^m-1}$ is
a morphism into $\mathfrak{M}_{2^m-1}$ and hence no sum of
$f_{2^m-1}(\bar{x}|_{2^m})$ for $\uparrow\bar{x}<2^m$ can give $\{1,
\ldots, 2^m\}$. Thus, it is necessary that $\bar{y}$ is above
the $\bar{x} \in At(n)$ s.t. $x_n = w_-$. This means, however,
that the $\bar{x} \in At(n)$ s.t. $x_n = w_+$ is not below
$\bar{y}$ and thus we obtain (by repeating  the previous argument)
that  $f_n(\bar{y}) \nsupseteq \{2^m+1, \ldots, 2^p\}$.
\item The forth condition --- $\bar{x}R_n\bar{y}$ implies
  $f_n(\bar{x}) \subseteq f_n(\bar{y})$ --- follows directly from
  definition of $f_n$; if $\bar{z} \in At(n)$ and  $\bar{z}R_n\bar{x}$
  then $\bar{z}R_n\bar{y}$.
\item That $f_n$ is onto and that it satisfies the back condition
  --- $f_n(\bar{x})\subseteq Y$ implies the existence of $\bar{z}$
  s.t. $\bar{x}R_n\bar{z}$ and $f_n(\bar{z}) = Y$ --- may be
  established as follows. Assume that $Y \nsupseteq \{1,
  \ldots, 2^m\}$. It means that $f_n(\bar{x}) \nsupseteq \{1,
  \ldots, 2^m\}$; let us denote by $\bar{u_1}$ the supremum of
$\bar{y} \in At(n)$ s.t. $\uparrow\bar{y} < 2^m$ and
$\bar{y}R_n\bar{x}$ (we use here the fact that for any element of
$\mathfrak{F}_n$, the set of all its precedessors forms a
lattice). Thus, $f_n(\bar{x}) \cap  \{1,
  \ldots, 2^m\} = f_n(\bar{u_1})$, $f_n(\bar{u_1}) = f_{2^m
  -1}(\bar{u_1}|_{2^m})$ and $f_n(\bar{u_1}) \subseteq Y \cap \{1,
  \ldots, 2^m\}$.  By the induction assumption that $f_{2^m-1}$
satisfies the back condition, there exists $\bar{v_1} \in
\mathfrak{F}_{2^m-1}$ s.t. $\bar{u_1}R_{2^m-1}\bar{v_1}$ and
$f(\bar{v_1}) = Y \cap \{1, \ldots, 2^m\}$. Now, if $Y \supseteq
\{2^m+1, \ldots, 2^p\}$, then $\bar{u_2}$ will be defined as the
supremum of all $\bar{y} \in At(n)$ s.t. $2^m \leq \uparrow\bar{y} <
n$ and $\bar{y}R_n\bar{x}$ and the $\bar{y} \in At(n)$
s.t. $\bar{y}_n = w_+$; then define $\bar{v_2} =
_{2^m-1}|\bar{u_2}$. Otherwise, $\bar{v_2}$ is constructed analogously as
$\bar{v_1}$ with the addition of $w_0$ in the last coordinate. Now,
concatenation of $\bar{v_1}$ and $\bar{v_2}$ is the desired $\bar{z}$. The
case when $Y \supseteq \{1,
  \ldots, 2^m\}$ but $Y \nsupseteq \{2^m+1,
  \ldots, 2^p\}$ is dealt with analogously.  \hfill $\dashv$
\end{itemize}

\begin{cor} \label{cor:proml}
{\sf Cheq} is a proper sublogic of {\sf ML}. Thus, it is also a
sublogic of {\sf KS} --- the logic of weak law of excluded middle.  {\sf Cheq} is not structurally complete.
\end{cor}

\noindent{\bf Proof}. The fact that {\sf Cheq} is a sublogic of {\sf ML}
follows from Theorem \ref{pmorph}. The fact that it is a proper sublogic
follows from Theorem \ref{kp}, as {\sf ML} contains the Kreisel-Putnam
axiom $\mathsf{kp}$. It may be proven also in a more direct way: the
Jankov formula of $\mathfrak{F}_2$ belongs to {\sf ML}, which may be
proven inductively, using self-resemblance property. However, this
would not give us any new information: Theorem \ref{kp}
implies that {\sf ML}$\nsubseteq$Log$\mathfrak{F}_2$ and
that is equivalent to the fact that the Jankov formula of
$\mathfrak{F}_2$ belongs to {\sf ML}.

The fact
that {\sf Cheq} cannot be structurally complete follows from Theorem
\ref{kp}, \ref{dis} and the recent result of Grigolia \cite{grig}, according to
which {\sf ML} is the only intermediate logic which is both
structurally complete and has the disjunction property. \hfill
$\dashv$

\vspace{0.2cm}

\noindent
Theorem \ref{pmorph} shows that there exists an interesting
semantical connection between {\sf Cheq} and {\sf ML}. A natural
question is then how close the connection is. Is it true, e.g., that
{\sf ML} is decidable iff {\sf Cheq} is? Or maybe {\sf ML} is finitely
axiomatizable over {\sf Cheq} (that would imply that {\sf Cheq} is
not axiomatizable by any set of formulas in finitely many
variables)?

For a while, the author hoped that {\sf Cheq}+{\sf kp} is a strong
sublogic of {\sf ML}, possibly even equal to {\sf ML}
itself. However, this hope was frustrated by the following
observation.

\begin{tm} \label{f2}
There exists a p-morphic image of ${\mathfrak F}_2$ which verifies
{\sf kp} but is not a p-morphic image of any $\mathfrak{M}_n$.
\end{tm}

\begin{figure}[hbtp]
  \begin{center}
    \begin{picture}(50,40)

      \multiput(10,20)(30,0){2}{\line(0,1){10}}
      
      \put(10,20){\line(3,-2){15}}
      \put(10,30){\line(2,-1){20}}

      \put(40,20){\line(-3,-2){15}}
      \put(40,30){\line(-2,-1){20}}

      \put(10,30){\line(1,-1){10}}
      \put(40,30){\line(-1,-1){10}}
  
      \put(25,10){\line(-1,2){5}}
      \put(25,10){\line(1,2){5}}

      \put(24,8){$r$}
      \put(7,19){$a$}
      \put(17,19){$b$}
      \put(32,19){$c$}
      \put(42,19){$d$}
      \put(7,29){$e$}
      \put(42,29){$f$}

    \end{picture}
    \caption{The frame $\mathfrak{H}$.}
    \label{fig:pm}
  \end{center}
\end{figure}
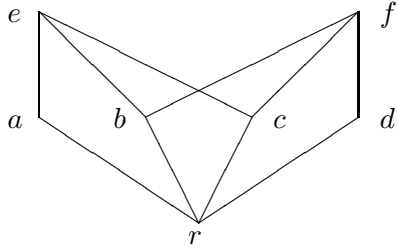

\noindent{\bf Proof}. Consider the frame $\mathfrak{H}$ depicted in Figure \ref{fig:pm}. We
will prove by induction that this is not a p-morphic image of
$\mathfrak{M}_n$. For $n=2$ it is obvious by a cardinality
argument. For $n=m+1$, by (IH) there is no p-morphism from any
proper generated subframe of $\mathfrak{M}_n$ onto $\mathfrak{H}$,
so the only set mapped onto $r$ is the empty
set, and there must be distinct singletons $\{x\}$ and $\{y\}$
mapped onto $a$ and $d$, respectively. The complement of $\{y\}$
(denoted by $Y$) and the
complement of $\{x\}$ (denoted by $X$) must then be mapped onto $e$ and
$f$,
respectively. It means then that $X \cap Y$ has to be mapped either
onto $b$ or $c$. Wlog assume it is mapped onto $b$. Then there has to be $Z
\nsubseteq X \cap Y$ mapped onto $c$ . However, it must be either the case
that $\{x\} \subseteq Z$ or $\{y\} \subseteq Z$. Assume $\{x\}
\subseteq Z$. Then, by applying the forth condition, the image of
$\{x\}$ (equal to $a$) is below $c$, a contradiction. \hfill $\dashv$

\vspace{0.2cm}

\noindent
Thus, to investigate further the connection between {\sf Cheq} and
{\sf ML} it is crucial to know
whether the techniques of Maksimova et al. \cite{maks} can be
applied in case of {\sf Cheq}. Let us recall that the
above-mentioned paper used a sequence of frames, which we
will denote here as $\{\mathfrak{G}_n\}_{n \geq 1}$ (in the original
paper they were denoted as $\{\Phi_k\}_{k \geq 1}$). It was proven
that (A) none of $\mathfrak{G}_n$ validates {\sf ML}, and hence the Jankov formula of $\mathfrak{G}_n$ belongs to {\sf ML} for every $n$; (B) that for
every $n$ and every $i \leq n$, a modification of $\mathfrak{G}_n$
denoted as $\mathfrak{G}^i_n$ validates {\sf ML} and hence canonical
formula of $\mathfrak{G}^i_n$ does not belong to {\sf ML}; (C) that
for a $n$-formula $\varphi$  (an $n$-formula is a formula
in $n$-variables), $\varphi$ entails the canonical formula of
$\mathfrak{G}_n$ only if it entails the canonical formula of
$\mathfrak{G}^i_n$ for some $i \leq n$. 

Now, the question is whether the frames from the sequence
$\{\mathfrak{G}_n\}_{n \geq 1}$ verify {\sf Cheq}. If so, it means
that {\sf ML} is not finitely axiomatizable over {\sf Cheq} and the
connection between the two systems is much weaker than Theorem
\ref{pmorph} would seem to suggest. If there are no suitable p-morphims, it
means that {\sf Cheq}, like {\sf ML} has no axiomatization in finitely many
variables. The author has not been able to generalize  Lemma 6 from
Maksimova et al. \cite{maks}, which was crucial for the main
result. On the other hand, he has not been able to define suitable
p-morphisms either. Anyway, the answer to this question would
make clear how similar (or how different) {\sf Cheq} and {\sf ML} are.

{\bf Acknowledgments}. The author wishes to thank Guram
Bezhanishvili for his attention, criticism and comments.





\noindent
\st{School of Information Science, JAIST} \\
\st{Asahidai, Tatsunokuchi, Ishikawa} \\
\st{923-1211 Japan} \\
\st{e-mail: \texttt{litak@jaist.ac.jp}}

\end{document}